\documentclass[draft,amsmath,amssymb,superscriptaddress,footinbib]{aipproc}
\layoutstyle{6x9}

\usepackage[english]{babel}
\usepackage{latexsym}
\usepackage{graphics}
\usepackage{subfigure}
\usepackage{epsfig}
\usepackage{amsfonts}
\usepackage{amssymb}
\usepackage{amsmath}
\usepackage{bbm}

\begin{document}

\title{Travelling to exotic places with ultracold atoms}



\author{M. Lewenstein}{
address={ICREA and ICFO--Institut de
Ci\`encies Fot\`oniques, E-08860 Castelldefels, Barcelona, Spain.}
}
\author{A. Kubasiak}{
address= {ICFO--Institut de Ci\`encies
Fot\`oniques, E-08860 Castelldefels, Barcelona, Spain.}
}
\author{J. Larson}{
address={ICFO--Institut de Ci\`encies
Fot\`oniques, E-08860 Castelldefels, Barcelona, Spain.}
}
\author{C. Menotti}{
address={ICFO--Institut de Ci\`encies
Fot\`oniques, E-08860 Castelldefels, Barcelona, Spain.},
altaddress={BEC-INFM-CNR, Physics Department, University 
of Trento, I-38100, Trento, Italy.}
}
\author{G. Morigi}{ 
address={Departament F\' isica, Grup d\'Optica,
Universitat Aut\'onoma de Barcelona, E-08193 Spain.}
}
\author{K. Osterloh}{
address={Institut f\"ur Theoretische Physik, Universit\"at
Hannover, D-30167 Hannover, Germany.} 
}
\author{A. Sanpera}{ 
address= {ICREA and Departament F\' isica, Grup de F\' isica Te\'orica, 
Universitat Aut\'onoma de Barcelona, E-08193 Spain.}
}

\begin{abstract}
We review and speculate on two recent developments of quantum optics
and ultracold atoms.  First, we discuss a possible realization of
``phonon'' physics, or as we call it {\it refracton} physics with
optical lattices. To this aim we combine the physics of cold atoms
with cavity QED, and investigate superfluid -- Mott insulator quantum
phase transition.
The systems can exhibit cavity mode modifications due to local changes
of refraction index ({\it refractons}).  Second, we discuss the
physics of strongly correlated particles in Abelian, and more
interestingly in non-Abelian magnetic fields, using cold atoms.
\end{abstract}

\maketitle

\section{Introduction}
 This paper has been presented as the final invited lecture at the
International Conference on Atomic Physics, ICAP 2006, held in
Innsbruck in July 2006.  The speaker, M. Lewenstein has met all the
advantages and disadvantages of being the last.  The audience began to
shrink and people were definitely starting to be tired of atomic
physics, despite the enormous success of the ICAP 2006. On the other
hand, being the last speaker M. Lewenstein had the opportunity to
speculate in his lecture upon the future of atomic physics, and the
physics of cold atoms in particular, and to thank truly the
organisers.

This paper consist of two relatively independent parts.  In the first
one we surmise the possibility of realising fascinating phenomena
resembling those occurring in solid state physics due to the presence
of phonons.  We start by reviewing the superfluid-Mott insulator
(SF--MI) transition of bosonic atoms in an optical lattice, which is a
paradigm of a quantum phase transition. We turn then to the SF--MI
transition inside an optical cavity.  In the simplest situation, the
atoms influence the cavity field by shifting the cavity resonance. The
resulting effective Bose-Hubbard model suggest the existence of
overlapping, competing Mott phases for wide range of parameters, and
bistable behavior in the vicinity of the shifted cavity resonance. In
a more sophisticated version, the model includes the possibility of
{\it local modifications} of the cavity field due to the presence of
atoms and local change of refractive index i.e. {\it refracton
physics}.

The second part deals with ultracold atomic gases in "artificial"
non-Abelian fields.  Again we review first distinct ways of realizing
Abelian and non-Abelian magnetic field in optical lattices using
various types of laser manipulations etc.  We present our recent
results dealing with the problem of a single atom in a lattice, which,
in the presence of Abelian (non-Abelian) magnetic fields, leads to a
spectrum in the form of the famous Hofstadter butterfly (or yet not so
famous {\it moth}). Finally, we speculate on non-Abelian fractional
Hall effect with ultracold atoms.

\section{Toward "refracton" physics}

\paragraph{\bf Introduction}
Ultracold atomic gases in optical lattices (OL) are nowadays a subject
of very intensive studies, since they provide an unprecedented
possibility to study numerous challenges of quantum many body physics
(for reviews see \cite{bloch,ml}). In particular, such systems allow
to realize various versions of Hubbard models \cite{zoller}, a
paradigm of which is the Bose-Hubbard model \cite{fisher}.  This model
exhibits supefluid (SF) -- Mott insulator (MI) quantum phase
transition \cite{sachdev}, and its atomic realization has been
proposed in the seminal Ref. \cite{jaksch}, followed by the seminal
experiments of Ref. \cite{blochex}. Several aspect and modifications
of SF -- MI quantum phase transition, or better to say crossover
\cite{batroumi}, has been intensively studied recently
(cf. \cite{ml,Bloch04}).  The control parameter of the SF -- MI
transition is the strength of the optical potentials, i.e. laser
intensity.  For low laser intensities, bosonic atoms can easily tunnel
around the lattice, and they condense into a maximally delocalized
states with long range phase coherence, and density (on--site atom
number) fluctuations. When laser intensity increases, optical
potential barriers prevent tunneling.  In effect, density fluctuations
become costly, and the systems enters a gaped, incompressible MI
phase, with no phase coherence, and reduced density fluctuations.

\begin{figure}[tbp]
\centering
\includegraphics[width=0.70\textwidth]{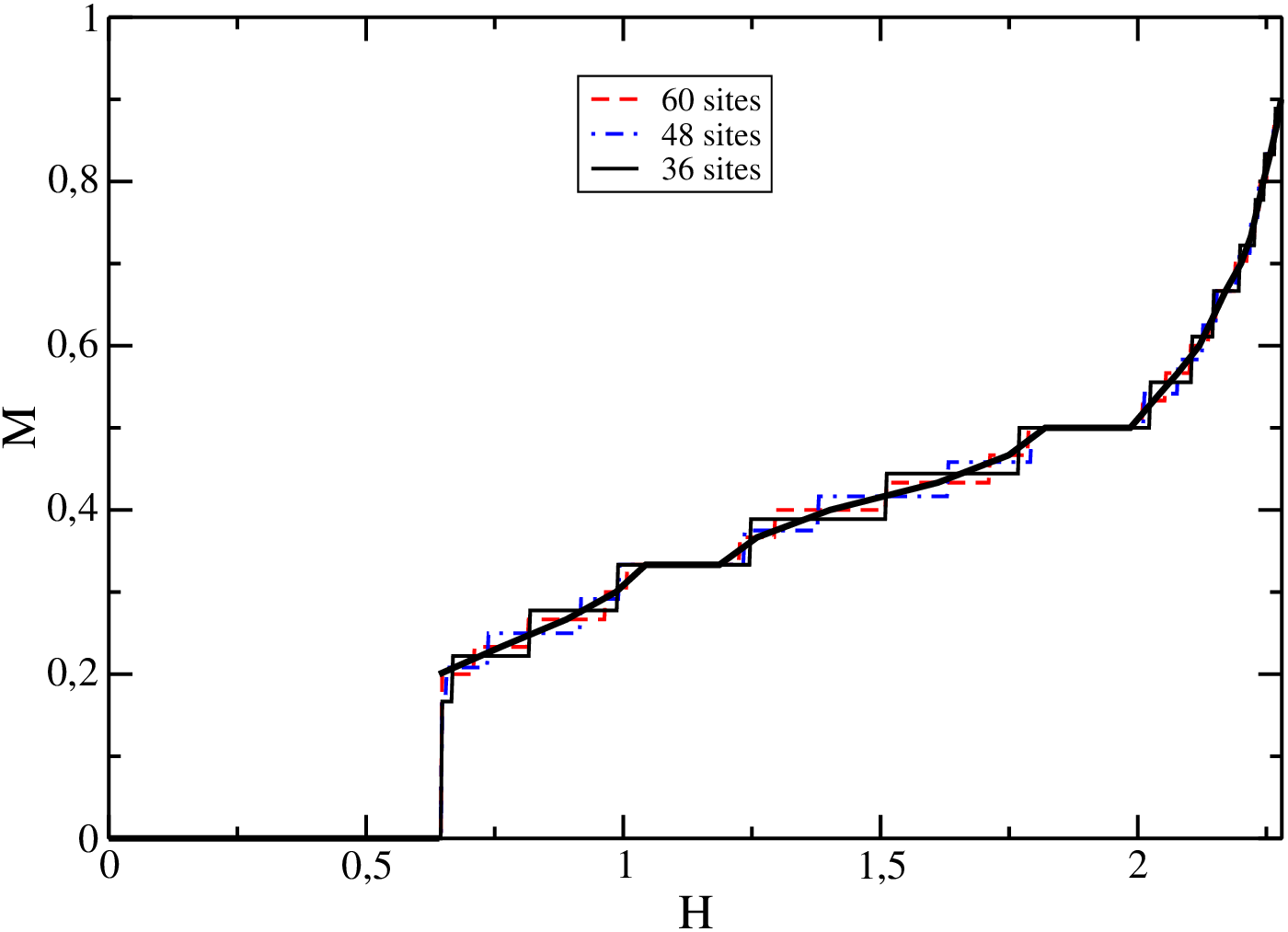}
\vspace*{0.5cm}
\caption{ Magnetization plateaux in a frustrated Heisenberg spin chain
analogous to the one described by expression (\ref{vekham}) (from
\cite{vekua}). }
\label{vekua-fig}
\end{figure}

\paragraph{\bf The critique of pure perfectionism}
Optical lattices provide an ideal toolkit for Hubbard models
\cite{zoller}, since they are robust and do not exhibit phonons -- in
this sense they are {\it perfect}. As we often encounter in everyday
life, perfectionism has positive and negative sides.  In particular,
the absence of phonons limits the whole richness of condensed matter
phenomena that result from electron-phonon interactions, such as for
instance the famous Peierls instability.  Antiferromagnetic
(non-frustrated) Heisenberg spin systems in perfect lattices, and in
particular spin chains in 1D at zero temperature often attain Ne\'el
ordering.  In the presence of phonons, the system arranges in such a
way that every second site move slightly to left, and every other
second to the right, so that the energy of spin interactions is
modulated: strong-weak-strong-....  The ground state has a form of a
chain of dimers (singlets) located at the "strong" bonds.  Contrary to
Ne\'el antiferromagnets which magnetize in arbitrarily small magnetic
fields $H$, a dimerized antiferromagnet requires a a non-zero
magnetization $M$ to break the dimers.  
Antiferromagnetic spin chains exhibit thus a magnetization plateau at
$M=0$.  Magnetization plateaux with much complex origin have been
recently intensively discussed in the literature (cf. \cite{vekua} and
references therein).  For instance, a simple XXZ spin chain, in an
adiabatic approximation for "heavy" phonons is described by the
Hamiltonian:
\begin{equation}
H=\frac{1}{2}\sum_i \delta_i^2 + J\sum_i(1-A\delta_i)(s^{\dag}_i
s_{i+1}+s^{\dag}_{i+1} s_{i}) + \Delta \sum_i
s_{z,i}s_{z,i+1},\label{vekham}
\end{equation}
where $\delta_i$ are lattice sites displacements, $J$ and $\Delta$
nearest neigbour XY and Ising spin couplings respectively, and $A$
describes the XY coupling modifications due to site displacements.
This system is expected to exhibit novel magnetization plateaux at
non-trivial rational values $M=1/3,1/2..$ (see
Fig. \ref{vekua-fig}). In the ground state, the displacement attains
the self-consistent value $\delta_i=\langle s^{\dag}_i
s_{i+1}+s^{\dag}_{i+1} s_{i}\rangle$. It is challenging and
fascinating to ask whether such physics can be captured and mimicked
by cold atoms in lattices that are not perfect and react to their
presence.  A natural candidate for the realization of such systems is
offered by Cavity Quantum Electrodynamics (CQED)~\cite{private}.

\paragraph{\bf Cavity QED and cold atoms}
In fact, in CQED the atoms interact with the cavity mode and actively
affect the cavity induced optical lattice. Several studies address the
scaling of cavity QED dynamics with the number of atoms and their
temperature. Cavity QED techniques were used to measure pair
correlations in atom laser \cite{esslinger}. Self-organization of
atoms in longitudinally and transversally pumped cavity was observed
in~\cite{black}, and theoretically characterized in~\cite{asboth}.
Bragg scattering of atomic structures, which built up in the potential
of the optical resonator, has been investigated in~\cite{Zimmermann}.
The interactions and cooling of a single atom in a quantum optical
lattice in a cavity was studied in Refs. \cite{domokosjosa,griessner}.
Maschler and Ritsch~\cite{Maschler} implemented the Bose-Hubbard model
for atoms in the one-dimensional potential of an optical resonator,
which was longitudinally pumped. They showed that the coefficients
determining the dynamics, depend on the number of atoms and exhibit
long range interactions, due to the photon-mediated forces.  In a
recent work, we have applied the model developed in~\cite{Maschler} to
study how the quantum phases of ultracold atomic gases are modified by
the photon-mediated long-range interaction due to the optical
resonator ~\cite{jonas}.  We focused onto the superfluid - Mott
insulator phase transition for gases of few hundred atoms, and apply
the methods of quantum statistical physics, namely strong coupling
expansion \cite{monien} to calculate the boundaries of the Mott
insulating phases in the $\mu-\eta$ plane, where $\mu$ is the chemical
potential, while $\eta$ characterizes the pump strength. We predicted
existence of overlapping, competing Mott phases, that may even consist
of two not connected regions for a wide range of parameters. We
predicted also dispersive bistable behavior \cite{carmichael} in the
vicinity of the shifted cavity resonance in the strong coupling
regime.

\paragraph{\bf Our model}
Our model considers atoms confined in one dimension inside an optical
resonator, which is pumped by a classical field.  The atomic dipole
transition is far-off resonance from the cavity mode, which induces a
dipole potential on the atomic ground state.  Using the notation
of~\cite{Maschler}, the single-particle Hamiltonian reads:
\begin{equation}\label{H:0}
H_0=\frac{p^2}{2m}+\hbar U_0\cos^2(kx)n_{\rm
ph}-\hbar\Delta_cn_{\rm ph}-i\hbar\eta\left(a-a^\dagger\right).
\end{equation} 
Here, $p$ and $m$ are the atomic momentum and mass, $\eta$ is the
amplitude of the pump at frequency $\omega_p$,
$\Delta_a=\omega_p-\omega_a$ and $\Delta_c=\omega_p-\omega_c$ are the
detunings of the pump from atom and cavity frequencies,
$g(x)=g_0\cos(kx)$ is the atom-cavity mode coupling, $k=\omega_c/c$
the mode wave vector, $a^\dagger$ and $a$ the creation and
annihilation operators of a photon $\hbar\omega_c$, $n_{\rm
ph}=a^{\dagger}a$ the number of photon, and $U_0=g_0^2/\Delta_a$ is
the depth of the single-photon dipole potential. The many-body
Hamiltonian is obtained from Eq.~(\ref{H:0}) in second quantization
with the atomic field operators $\Psi(x)$, and including the atomic
contact interaction~\cite{Maschler}. In the bad-cavity limit we
eliminate, from the atomic dynamics, the cavity field variables and
the number of photons takes the value
\begin{equation} n_{\rm
ph}\approx \frac{\eta^2}{\kappa^2+\left[\Delta_c-U_0\int {\rm
d}x\cos^2(kx)\Psi^{\dagger}(x)\Psi(x)\right]^2}, \end{equation} 
with
$\kappa$ the rate of cavity damping. The cavity potential depends thus
non-linearly on the atomic density. In particular, certain atomic
phases correspond to resonances, which increase the number of photons
and thus the optical lattice depth. This is the salient physical
property of this system, which gives rise to novel dynamics, as we
will show. We now consider the regime of validity of the tight binding
approximation (TBA) and expand the atomic field operators in the
lowest energy band as $\Psi(x)=\sum_ib_iw(x-x_i)$, where $b_i$ is the
atomic annihilation operator at site $i$, and $w(x-x_i)$ is the
Wannier state localized around site $i$, which depends on the photon
number $n_{\rm ph}$, and thus on the atomic density. In the Wannier
expansion we keep on-site and nearest-neighbour couplings, and neglect
all other couplings, and obtain the Hamiltonian
\begin{equation}\label{effham3} H/U =
-tB+\frac{1}{2}\sum_in_i(n_i-1)-g B^2-\mu N, \end{equation} 
where
$N=\sum_in_i=\sum_ib_i^\dagger b_i$ is the atom number operator,
$B=\sum_i b_i^\dagger b_{i-1}+\mathrm{h.c.}$ is the hopping term,
and $U$ denotes the strength of on--site interactions. The
parameters 
\begin{equation} t=-\frac{E_{1}}{U}+\frac{\eta^2\hbar
U_0J_{1}\left(\kappa^2-\zeta^2\right)}{U\left(\kappa^2+\zeta^2\right)^2};\;\;
g=-\frac{\eta^2\hbar
U_0^2J_{1}^2\zeta\left(3\kappa^2-\zeta^2\right)}{U\left(\kappa^2+\zeta^2\right)^3},
\end{equation} 
denote tunneling and long-range coupling respectively, and are
expressed in terms of the $N$-dependent term $\zeta=\Delta_c-U_0J_0N$
and of the integrals $E_{\ell}=\int
dx\,w(x-x_l)(-{\hbar^2}/{2m})({d^2}/{dx^2})w(x-x_{l+\ell})$ and
$J_{\ell}=\int dx\,w(x-x_l)\cos^2(kx)w(x-x_{l+\ell})$, with
$\ell=0,1$, where we have assumed $J_1\ll J_0$. Note that in
Eq.~(\ref{effham3}) the number of particles is conserved, since
$[N,H]=0$.

\paragraph{\bf SF - MI transition in  a cavity}

\begin{figure}[th] \includegraphics[width=0.8\linewidth]{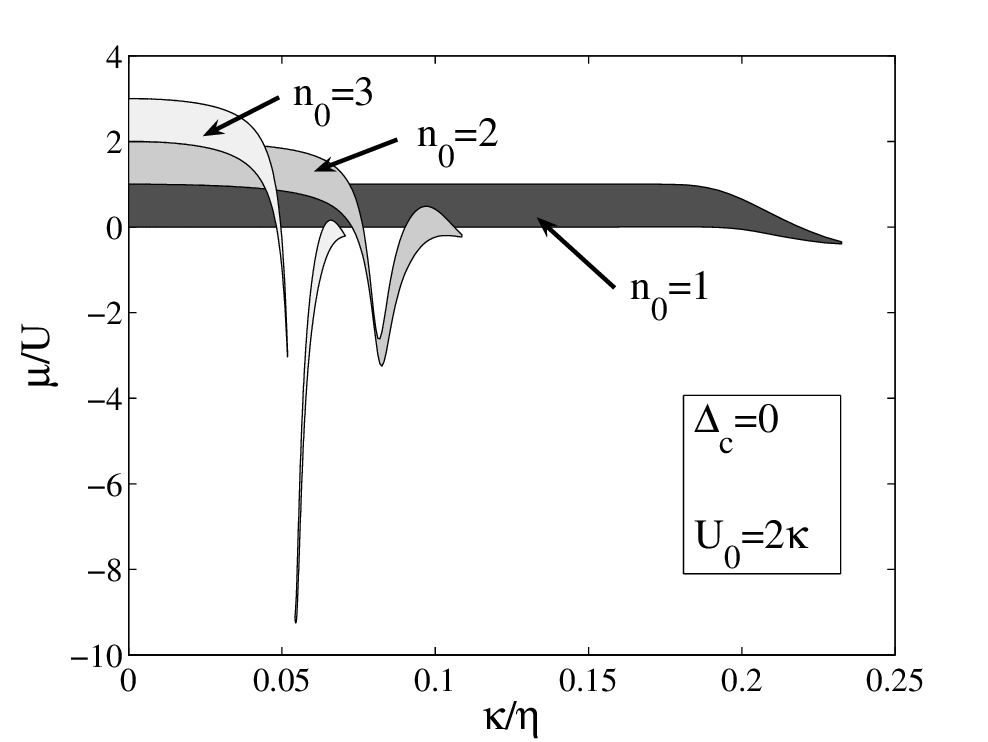}
\caption{Overlapping and disconnected Mott insulator regions for
$\Delta_c=0$ and $U_0=2\kappa$.}
\label{fig2} \end{figure}

In \cite{jonas} we have generalized and applied the method of
H. Monien's group \cite{monien} (for more references see \cite{ml})
to calculate the lobes for a modified 1D Bose-Hubbard model
(\ref{effham3}) describing atoms in an optical lattice created by
pumping a laser beam into the cavity. Technically, this corresponds
of using perturbation theory in $B$ to calculate the ground state
energy, and the energies of excited states with one additional atom
(one hole) present and comparing them.  The major difference to the
standard case is that the intensity of the cavity field depends on
the number of atoms present, since the atoms shift collectively the
cavity resonance. Thus the coefficients $t$, $U$, $g$ become very
complicated functions of the cavity detuning, intensity of the
pumping laser, $N$, etc. Moreover, quantum fluctuations of the
resonance shift induce long range interactions between the atoms,
proportional to $g$ and hence the parameters $t$, $U$, $g$ have to be
calculated self-consistently. Moreover these calculations sometimes
exhibit bistability effect!
Instead of using $\mu$ and $t$ to describe the phase diagram, it is
necessary to use a proper control parameter which is the strength of
the pumping laser.  A typical phase diagram as a function of $\mu$ in
the recoil units, and $\kappa/\eta$ is shown in Fig. \ref{fig2}.  The
striking effect is the overlap a different Mott phases, and even a
presence disconnected Mott regions, which follows from the fact that
the expressions for $t$ and $U$ for $n=1, 2,3, \ldots$ Mott phases are
different.

Finally, if $\Delta_a$ and $\Delta_c$ have the same sign, and one may encounter resonances such that
$\Delta_c-U_0J_0N=0$; the system may exhibit a bistable behaviour, 
as reported in \cite{jonas}

\paragraph{\bf Toward genuine "refractons"}
An obvious challenging question is whether one can generalize the
above results to a situation when the atoms affect the cavity mode
{\it locally}. Such a regime of CQED is experimentally feasible and
has been recently considered by D. Meiser and P. Meystre
\cite{pierre}, who termed it "the Super-Strong Coupling Regime of
Cavity Quantum Electrodynamics".  We present here a somewhat
alternative approach based on the techniques that one of us developed
together with R. Glauber, T. Mossberg and K. Zakrzewski to describe
spontaneous emission and Lamb shift modifications in dielectric media
and/or in "colored" photonic reservoirs \cite{glauber}. The idea is to
use the full quantum electrodynamical description of the electric
field inside the cavity, i.e. expand it into the continuum of modes
that have a significant maximum in the density of modes at the cavity
resonance. The single atom Hamiltonian becomes then
\begin{eqnarray}
H_0&=&\frac{p^2}{2m}+\hbar U_0\hat E^{\dag}(x)\hat E(x) + \int dk\,
ck\, \hat a^{\dag}(k)\hat a(k)\nonumber\\ &-&i\hbar\eta\int dk\,p(k)
\left(\hat a(k) \exp{(-ic|k|t)}-\hat a^{\dag}(k) \exp{(+ic|k|t)}
\right)\cos{(kx)},\label{ham4}
\end{eqnarray} 
where $\hat a(k)$, $\hat a^{\dag}(k)$ are photon annihilation and
creation operators of photons in standing wave modes (that inside the
cavity may be well described by $\cos(kx)$).  The pumping laser pulse
is expanded into these modes, and the amplitude of expansion, $p(k)$
is strongly peaked at $k\simeq k_L$, and $\omega=c|k|\simeq
c|k_L|=\omega_L$.  The electric field operator $\hat E(x)$ inside the
cavity can be expanded as (for simplicity we consider only even
modes):
 \begin{equation}\label{e}
\hat E(x) \propto \int dk\, \frac{\kappa}{(k-\omega_c+i\kappa)}\hat
a(k)\cos(kx)=\frac{\eta \cos(k_Lx)}{\kappa - i
(\Delta_c-U_0\hat\delta(x))},
\end{equation} 
where the final expression appears after adiabatic elimination of the
cavity field.  The AC Stark shift (which was given by $\delta=J_0\hat
N + J_1 \hat B$) becomes now {\it local}, i.e. $x_i$-dependent:
\begin{equation}
\hat\delta(x_i)= J_0\sum_j 1/2 \left(\exp(-\kappa|x_i-x_j|) +
\exp(-\kappa|x_i+x_j|\right)\hat n_j + J_1 \hat B
\end{equation}
with $\hat n_j$ being the atom number operator at the site $j$.  

In the hard boson limit we end up with the Hamiltonian: 
\begin{eqnarray}
H=&-&\sum_i \left[J(i, \{s_{z,j}\})s^{\dag}_i s_{i+1}+s^{\dag}_{i+1}
s_{i} J(i,\{s_{z,j}\})\right]\nonumber\\ &-&\sum_i \left[\Delta(i,
\{s_{z,j}\})s^{\dag}_i s_{i}+s^{\dag}_{i} s_{i} \Delta(i
\{s_{z,j}\})\right] - \sum_i H s_{z,i},\label{vekham-lattice}
\end{eqnarray}
Apart from the fact that the system is ferromagnetic, it resembles to
great extend the spin chain with adiabatic phonons (\ref{vekham}): the
main difference being that the self-consistent hopping amplitudes
depend functionally on the density ($n_j$'s, or equivalently
$s_{z,j}$'s), rather than on spin-spin correlations ($\langle
s^{\dag}_i s_{i+1}+s^{\dag}_{i+1} s_{i}\rangle$.

\section{\bf Ultracold gases in ``artificial'' non-Abelian fields}
\paragraph{\bf Ultracod atoms and HEP} 
In the recent years the links and interconnections between physics of
ultracold atoms and condensed matter physics became solid and well
established. Through these links, modern atomic physics reaches the
frontiers of the quantum field theory, and has indirect relations to
the modern high energy physics.  Very recently, it has appeared
several proposals on how to use ultracold atoms to simulate abelian
$U(1)$ lattice gauge theory \cite{lgt}.  Another research line
investigates the possibility of inserting atoms in artificial
non-Abelian magnetic fields by employing electromagnetically induced
transparence \cite{patrick}, or optical lattices \cite{klaus}. The
recent progress in the latter case will be presented here.

 \paragraph{\bf Artificial Abelian "magnetic fields" in a lattice}
As it is well known, rapidly rotating harmonically trapped gases of
neutral atoms exhibit effects analogous to charged particles in
uniform magnetic fields (for a recent overview, see \cite{nuri}).
Thus one should be able to realize analogues of fractional quantum
Hall effect (FQHE) in such systems.  However, to achieve a regime of
``strong'' magnetic fields it is desirable to realize artificial
fields in an optical lattice by controlling e.g. the phases of the
hopping amplitudes.  In particular, single particle stationary
Schr\"odinger equation in 2D square lattice in perpendicular magnetic
field, in Landau gauge (i.e. with the vector potential ${\bf
A}=(0,Bx,0)$) reads:
\begin{eqnarray}
E\Psi(m,n)&=&-J\left[\Psi(m+1,n)+ \Psi(m-1,n)+ 
\exp(-i2\pi\alpha m)\Psi(m,n+1)\right. \nonumber\\
&+&\left. \exp(+i 2\pi \alpha m)\Psi(m,n-1)\right],\label{2dabel}
\end{eqnarray}
where $(m,n)$ are integers enumerating coordinates on the lattice, and
$\alpha= Ba^2/2\pi(\hbar c/e)$, is the magnetic flux thought an
elementary plaquette of area $a^2$ in units of quantum flux. Assuming
plane waves in the $y$ direction, $\Psi(m,n)=\exp(i\nu n)g(m)$, we
obtain the famous Harper's equation:
\begin{equation}
\epsilon g(m)=-\left[g(m+1)+ g(m-1)+ 2\cos(2\pi \alpha
m-\nu)g(m)\right]\label{2dharper}.
\end{equation}

Jaksch and Zoller \cite{butterfly} were the first to propose how to
realize the model of Eq. (\ref{2dabel}) employing internal states of
atoms, and using appropriate combination of laser assisted tunneling,
lattice tilting, resonance pumping, etc.  They argued that such
systems could be used to study the Hofstadter butterfly (i.e. spectrum
of states) in the presence of weak atom-atom interactions. Several
other groups \cite{Mueller,Demler} proposed since then alternative
methods of realizing Abelian fields, aiming at the possibility of
realizing FQHE in such systems: Laughlin states at low $\alpha$
\cite{Demler}, and analogues of bilayer FQHE (cf. \cite{das}) Halperin
states \cite{Jaksch}.

\begin{figure}[th] \includegraphics[width=0.8\linewidth]{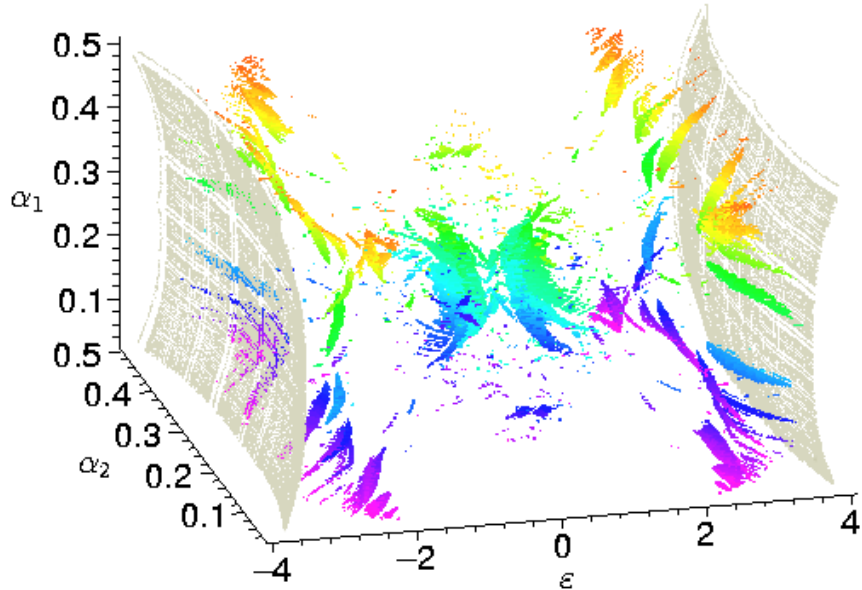}
\caption{Hofstadter moth's energy gaps plotted in various colours
against rational values of "magnetic fluxes" $\alpha_1$ and
$\alpha_2$.  Whole spectrum is contained between the two external,
slightly curved, grey "walls".}
\label{moth} \end{figure}

\paragraph{\bf Artificial non-Abelian "magnetic fields" in a lattice}
In \cite{klaus} we demonstrated that using atoms with multiple
internal states ("colours" or "lavours"), one can realize Non-Abelian
magnetic fields in an lattice using a similar scheme as
\cite{butterfly}, i.e.  employing laser assisted tunneling, lattice
tilting, resonance pumping etc. Generally speaking, the proposed
scheme allows, among other things, to generate fields, whose vector
potential is linear in $x$ and $y$, i.e.  is of the form ${\bf A}=
(\hat M_x + \hat N_x (x/a)+ \hat O_x (y/a), (\hat M_y + \hat N_y
(x/a)+ \hat O_y(y/a), 0)$, where $\hat M, \hat N$, and $\hat O$ are
arbitrary hermitian matrices from the algebra of the considered group,
which can be chosen practically at will: $SU(n)$, $U(n)$, $GL(n)$
etc. For instance, with ${\bf A}= (\hat M_x, \hat N_y (x/a), 0)$ with
$\hat M_x, N_y$ - hermitian, and $N_y=2\pi {\rm
diag}(\alpha_1,\alpha_2,\ldots,\alpha_N)$, we obtain the $U(n)$
generalization of the Harper's equation for the $n$-component wave
function:
\begin{equation}
\epsilon g(m)=-\left[\exp(i\hat M_x)g(m+1)+ \exp(-iM_x)g(m-1)+
2\cos(2\pi \hat N_y m-\nu)g(m)\right]\label{2dharpernon}.
\end{equation}
In Ref. \cite{klaus} we have studied the spectrum of this equation for
the $SU(2)$ case, and rational $\alpha_1$ and $\alpha_2$.  Bloch
theory applies in such a case and the spectrum consists of allowed
bands and gaps, similarly as in the Hofstadter problem for $n=1$ and
$\hat M_x=0$. In the latter case the allowed energy band plotted
against (rational) $\alpha$, form the complex fractal figure known as
Hofstadter butterfly. In the present case, we plot instead energy
gaps.  They form a fractal arrangements of genuine holes is the
spectrum (Fig. \ref{moth}), which we term Hostadter moth.

\begin{figure}[th]
\includegraphics[width=0.6\linewidth]{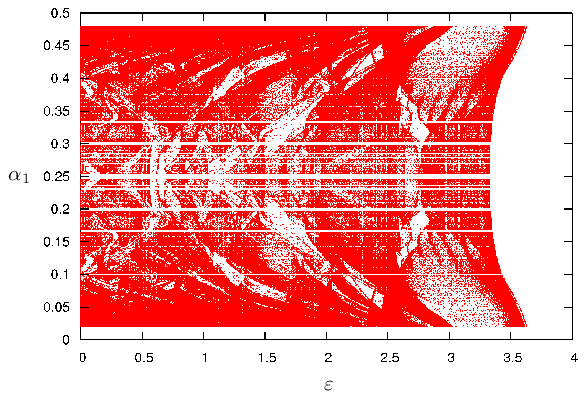}
\caption{Section of the Hofstadter moth at $\alpha_2=2/5$ in a lattice
of 1000 sites.}
\label{monster} 
\end{figure}

Despite its richness, Fig. \ref{moth} contain relatively little
informations on density of states and corresponds to an infinite
system.  We consider realistic systems of 1000 sites size, and plot a
section of the moth for a fixed value of, say, $\alpha_2=2/5$.  Such a
plot in which the red points (holes) indicate the presence (absence)
of an energy eigenvalue is, perhaps, the simplest representation of
the density of states (see Fig. \ref{monster}).  The moth section is
especially impressive if one turns the figure 90 degrees clockwise.
The moth looks like a horrible monster from a horrible Japanese
cartoons, that are being shown to the kids all over the world in order
to educate them in the spirit of war, killings, anger, etc.

\paragraph{\bf Non-Abelian fractional quantum Hall effect}  
We term by non-Abelian FQHE the analog of the "standard" (if one dares
to use such adjective) FQHE, but in the constant non-Abelian magnetic
fields. Our starting point is a non-abelian $SU(n)$ field in a 2D
square lattice. Note, that we cannot use the gauge potential from the
previous subsection, since it leads to a non-zero gauge field
tensor. Instead we use ${\bf A}=(-B(y-\tilde b\hat S_y)/2, B(x+\tilde
b\hat S_y)/2,0)$, where $\hat S_{x,y,z}$ are the spin operators
($S=(n-1)/2$). This potential has a non-vanishing constant rotation,
and is non-trivially non-Abelian, since its components do not
commute. We consider now not too strong fields so that the continuum
approximation in the lattice maybe performed.  The single particle
Hamiltonian becomes thus for $n=2$
\begin{equation}
H=\frac{(\hat p_x+eB(y-b\hat \sigma_y)/2c)^2}{2m} +\frac{(\hat
p_y-eB(x + b\hat \sigma_x)/2c)^2 }{2m} + \omega \hat\sigma_z/2,
\end{equation}
where we have included $\sigma_z$ term to allow for non-trivial
coupling of different Landau levels. Introducing the operators that
annihilate or create Landau level number, $\hat D=(\hat Q+i\hat
P)\sqrt{2}$, $D^{\dag}$, with $\hat Q=\sqrt{c/eB}(\hat p_x+eBy)$,
$\hat P=\sqrt{c/eB}(\hat p_y-eBx)$, the Hamiltonian attains the form
of the Jaynes-Cummings model \cite{carmichael}:
\begin{equation}
H/\hbar\Omega = \left[\hat D^{\dag}\hat D- i\beta
(\hat\sigma^{\dag}\hat D- \hat
D^{\dag}\hat\sigma)\right]+\tilde\omega\hat\sigma^{\dag} \hat\sigma,
\end{equation}
where $\Omega=eB/\hbar mc$, $\tilde\omega=\omega/\Omega$, and
$\beta=b\sqrt{eB/2c}$. The many body Hamiltonian includes the sum of
single particle terms plus the sum of pair interactions of contact
type,
\begin{equation}
{\cal H}= \sum_i\left[\hat D^{\dag}_i\hat D_i- i\beta
(\hat\sigma_i^{\dag}\hat D_i- \hat
D_i^{\dag}\hat\sigma_i)\right]+\tilde\omega\hat\sigma_i^{\dag}
\hat\sigma_i + \frac{g}{2}\sum_{i \ne j}\delta^{(2)}(z_i-z_j),
\label{hamany}
\end{equation} 
where $z_i=x_i+iy_i$ are complex coordinates in the plane.

There are some similarities between (\ref{hamany}) and a
multicomponent quantum Hall system studied in the literature (see
article by S.M. Girvin and A.H. Monald in \cite{das}), such as
electronic systems with spin, or bilayer systems.  In the spin analogy
our model introduces several novel aspects: i) we may use it for
fermions or bosons, ii) we may consider arbitrary spin (pseudo-spin),
and in particular consider contact interactions of the spinor type
(cf. \cite{ml}). Concerning multilayer analogy we stress that,
similarly, we may consider more general "multilayer"$=$multicolour
situations, and control coherent coupling between the "layers". Our
model breaks the spin up-down symmetry, and necessarily mixes various
Landau levels, if we allow for spin-up excitations.  If all spins are
down, the ground state (GS) takes the Lauglin form:
\begin{equation}
|\Omega,
 E=0\rangle=\prod_{i<j}(z_i-z_j)^m\exp(-\sum_i|z_i|^2/4)|\downarrow,
 \downarrow,\ldots,\downarrow\rangle,
\end{equation}
where $1/m=\nu$ is the filling factor, $\hat E=\sum_i \hat
D^{\dag}_i\hat D_i$ is the total number of excitations, and $E$ its
mean value.  Correspondingly, the GS with one excitation will be
\begin{equation}
|\Omega,
|E=1\rangle\rangle=\prod_{i<j}(z_i-z_j)^m\exp(-\sum_i|z_i|^2/4)|W(\uparrow)
\rangle
|+ \sum_i D^{\dag}_i \Omega, E=1\rangle
\end{equation}
It is natural to generalize the concept of hole excitations for the
non-Abelian FQHE, defining single hole states $\prod_i(z_i -
Z_a)|\omega, E\rangle$, two hole states $\prod_i(z_i -
Z_a)(z_i-Z_b)|\omega, E\rangle$ etc., where $Z_a$, $Z_b$ are $n\times
n$ matrices.  These excitations constitute obviously fractional
anyons, but currently we are working on checking whether they
themselves are non-Abelian, i.e. transform according the non-Abelian
representations of the permutation group \cite{read}). This has been a
long lasting quest for non-Abelian anyons in the condensed matter
literature, but without no clear experimental observation so far.  The
most prominent candidate is electronic (fermionic) $\nu=5/2$ FQHE
state.  This state has been observed in experiments, Moore and Read
\cite{moore}, and independently Greiter, Wen and Wilczek
\cite{greiter} proposed to explain it in terms of the "Pfaffian state"
(see also \cite{das}).  Recently, however, T\"oke and Jain \cite{toke}
proposed an alternative "composite fermions" model which does not
relate to non-Abelian statistics in any obvious manner. For bosons, a
promising candidate is the $\nu=3/2$ state from the so called
Read-Rezayi sequence of incompressible correlated liquids.  This state
seem to be a true ground state of the rapidly rotating gas of bosons
interacting via contact (Van der Waals) forces with a moderate amount
of dipolar interactions \cite{rezayi}. Such situation may be achieved,
for instance, with Bose condensed Chromium, as in experiments of the
T. Pfau's group \cite{chromium}. We hope that non-Abelian FQHE, due
its profound and direct non-commutative character, will provide
further, experimentally feasible examples of non-Abelian anyons.

We conclude with the standard conclusion of M. Lewenstein and the
motto to the review \cite{ml}, which is a citation from William
Shakespeare's "Hamlet": {\it There are more things in heaven and
earth, Horatio, than are dreamed of in your philosophy}. In the
present context, it expresses our neverending curiosity, enthusiasm,
joy, and excitement of working in atomic physics in general, and
physics the of ultracold atoms in particular.\\

We thank B. Damski, E. Demler and H. Ritsch for discussions.  We
acknowledge support from German Government(SFB 407, SPP 1116, GK 282,
436 POL), Spanish Government (FIS2005-04627,01369).  EU IP Programme
"SCALA", ESF PESC QUDEDIS.  G.M, C. M and J.L are supported by Spanish
Ramon y Cajal Program, EC-Marie Curie fellowship and Swedish
Goverment/Vetenkasprader respectively.


\begin{thebibliography}{999}

\bibitem{bloch} I. Bloch and M. Greiner,
Adv. At. Molec. Opt. Phys. {\bf 52}, 1 (2005).

\bibitem{ml} M. Lewenstein, A. Sanpera, V. Ahufinger, B. Damski,
A. Sen (DE), and U. Sen, cond-mat/0606771.

\bibitem{zoller} D. Jaksch and P. Zoller, Ann. Phys. (N.Y.) {\bf 315},
52 (2005).

\bibitem{fisher} M.P.A. Fisher, P.B. Weichman, G. Grinstein, and D.S.
Fisher, Phys. Rev. B {\bf 40}, 546 (1989).

\bibitem{sachdev} S. Sachdev, \textit{Quantum Phase Transitions},
(Cambridge University Press, Cambridge, 1999).

\bibitem{jaksch} D. Jaksch, C. Bruder, J.I. Cirac, C.W. Gardiner, and
P. Zoller, Phys. Rev. Lett. {\bf 81}, 3108 (1998).

\bibitem{blochex} M. Greiner, O. Mandel, T. Esslinger, T. W. Hänsch
and I. Bloch, Nature {\bf 415}, 39 (2002).

\bibitem{batroumi} G.G. Batrouni, V. Rousseau, R.T. Scalettar,
M. Rigol, A. Muramatsu, P.J.H. Denteneer, and M. Troyer, Phys. Rev.
Lett. {\bf 89}, 117203 (2002); G.G. Batrouni, F.F. Assaad,
R.T. Scalettar, and P.J.H. Denteneer, Phys. Rev. A {\bf 72}, 031601(R)
(2005).

\bibitem{Bloch04} I. Bloch, Physics World {\bf 17}, 25 (2004).

\bibitem{vekua} cf. T. Vekua, D.C. Cabra, A. Dobry, C. Gazza, and
D. Poilblanc, Phys. Rev. Lett. {\bf 96}, 117205 (2006), and references
therein.

\bibitem{private} B. Nagorny, T. Els\"asser, and A. Hemmerich,
Phys. Rev.  Lett. {\bf 91}, 153003 (2003); D. Kruse, C. von Cube,
C. Zimmermann, and Ph. W. Courteille, Phys. Rev.  Lett. {\bf 91},
183601 (2003); J. A. Sauer, K. M. Fortier, M. S. Chang, C. D. Hamley,
and M. S. Chapman Phys. Rev. A {\bf 69}, 051804 (2004); also E. Hinds,
J. Reichel, G. Rempe, A. Hemmerich, T. Esslinger, privite
communications.

\bibitem{esslinger} A. \"Ottl, S. Ritter, M. K\"ohl, and T. Esslinger
Phys. Rev. Lett. {\bf 95}, 090404 (2005).
\bibitem{black} A.T. Black, H.W. Chan, and V. Vuleti\'c, Phys. Rev. Lett.  {\bf 91}, 203001 (2003).

\bibitem{asboth} P. Domokos and H. Ritsch, Phys. Rev.  Lett. {\bf 89},
253003 (2002); J.K. Asb\'oth, P. Domokos, H. Ritsch and A. Vukics,
Phys. Rev. A {\bf 72}, 053417 (2005).

\bibitem{Zimmermann} S. Slama, C. von Cube, B. Deh, A. Ludewig,
C. Zimmermann, and Ph. W. Courteille, Phys. Rev. Lett. {\bf 94},
193901 (2005).


\bibitem{domokosjosa} P. Domokos and H. Ritsch, J. Opt. Soc. Am. B
{\bf 20}, 1098 (2003).

\bibitem{griessner} A. Griessner, D. Jaksch, and P. Zoller, J. Phys. B
{\bf 37}, 1419 (2004).

\bibitem{Maschler} C. Maschler and H. Ritsch, Phys. Rev. Lett.  {\bf
95}, 260401 (2005).

\bibitem{jonas} J. Larson, B. Damski, G. Morigi, and M. Lewenstein,
cond-mat/0608335.

\bibitem{monien} J.K. Freericks and H. Monien, Europhys. Lett. {\bf
26}, 545 (1994).

\bibitem{carmichael} cf. D.F. Walls and G.J. Milburn, {\it Quantum
Optics}, (Springer, Berlin, 2006).

\bibitem{pierre} D. Meiser and P. Meystre, quant-ph/0605020.

\bibitem{glauber} R.J. Glauber and M. Lewenstein, Phys. Rev. A {\bf
43}, 467 (1991); M. Lewenstein, T.W. Mossberg, and R.J. Glauber,
Phys. Rev. Lett. {\bf 59}, 775 (1987); M. Lewenstein, J. Zakrzewski,
and T.W. Mossberg, Phys. Rev. A {\bf 38}, 1075 (1988).

\bibitem{lgt} H.-P. B\"uchler, M. Hermele, S.D. Huber, M.P.A. Fisher,
and P. Zoller, Phys. Rev. Lett. {\bf 95}, 040402 (2005); J.K. Pachos
and E. Rico, Phys. Rev. A {\bf 70}, 053620 (2004); S. Tewari,
V.W. Scarola, T. Senthil, and S. Das Sarma, cond-mat/0605154.


\bibitem{patrick} J. Ruseckas, G. Juzeliunas, P. \"Ohberg, and
M. Fleischhauer, Phys. Rev. Lett.  {\bf 95}, 010404 (2005).

\bibitem{klaus} K. Osterloh, M. Baig, L. Santos, P. Zoller, and
M. Lewenstein, Phys. Rev. Lett.  {\bf 95}, 010403 (2005).

\bibitem{nuri} N.Barber\'an, M.Lewenstein, K.Osterloh, and D.Dagnino,
Phys. Rev. A {\bf 73}, 063623 (2006), cond-mat/0603200.

\bibitem{butterfly} D. Jaksch and P. Zoller, New J. Phys. {\bf 5},
Art. No. 56 (2003).
 
\bibitem{Mueller} E.J. Mueller, Phys. Rev. A {\bf 70}, 041603 (2004).

\bibitem{Demler} A.S. S\o rensen, E. Demler, and M.D. Lukin,
Phys. Rev. Lett. {\bf 94}, 086803 (2005).

\bibitem{das} S. Das Sarma and A. Pinchuk (Eds.), {\it Perspectives in
Quantum Hall Effects}, (Wiley \& Sons, New York, 1997)

\bibitem{Jaksch} R.N. Palmer and D. Jaksch, Phys. Rev. Lett. {\bf 96},
180407 (2006), cond-mat/0604600.

\bibitem{read} N. Read and G. Moore, hep-th/9202001.

\bibitem{moore} G. Moore and N. Read, Nucl. Phys. B {\bf 360}, 362 (1991).

\bibitem{greiter} M. Greiter, X.G. Wen, and F. Wilczek,
Phys. Rev. Lett. {\bf 66}, 3205 (1991); Nucl. Phys. B {\bf 374}, 567
(1992); Phys. Rev. B {\bf 46}, 9686 (1992).

\bibitem{toke} C. T\"oke and J.K. Jain, Phys. Rev. Lett. {\bf 96},
246805 (2006), cond-mat/0606610.

\bibitem{rezayi} E. H. Rezayi, N. Read, and N.R. Cooper,
Phys. Rev. Lett. {\bf 95}, 160404 (2005).

\bibitem{chromium} A. Griesmaier, J. Werner, S. Hensler, J. Stuhler,
and T. Pfau, Phys. Rev. Lett. {\bf 94}, 160401 (2005).




\end{thebibliography}
\end{document}